\documentclass[preprint,11pt]{elsarticle}

\usepackage{hyperref}
\usepackage[margin=2.5cm, top=2cm, bottom=2cm]{geometry}
\usepackage[sfdefault]{libertine}

\journal{Journal of \LaTeX\ Templates}

\biboptions{authoryear}

\usepackage{xcolor}
\usepackage{ulem}
\newcount\Comments  
\newcommand{\kibitz}[2]{\ifnum\Comments=1\textcolor{#1}{#2}\fi}

\newcommand{\blue}[1]{#1} 

\newcommand{\colorA}[1]{#1} 
\newcommand{\colorB}[1]{#1}
\usepackage{changepage}
\usepackage{array}
\newcolumntype{+}{!{\vrule width 2pt}}
\newlength\savedwidth

\newcommand\thickhline{\noalign{\global\savedwidth\arrayrulewidth\global\arrayrulewidth 2pt}%
\hline
\noalign{\global\arrayrulewidth\savedwidth}}

\makeatletter
\def\ps@pprintTitle{%
    \let\@oddhead\@empty
    \let\@evenhead\@empty
    \def\@oddfoot{\footnotesize\itshape
         {Submitted preprint. Please contact the authors before citing.} \hfill}%
    \let\@evenfoot\@oddfoot
    }
\makeatother

\begin{document}

\begin{frontmatter}

\title{A physiology-inspired framework for holistic city simulations}

\author[BSC,UIC]{Irene Meta\corref{mycorrespondingauthor}}
\cortext[mycorrespondingauthor]{Corresponding author. Barcelona Supercomputing Center (BSC), CASE Department, C/Jordi Girona 29, Nexus II Building, Barcelona, Spain}
\ead{irene.meta@bsc.es}

\author[BSC]{Fernando M. Cucchietti}
\author[UIC]{Diego Navarro}
\author[UDD]{Eduardo Graells-Garrido}
\author[IAAC]{Vicente Guallart}

\address[BSC]{Barcelona Supercomputing Center (BSC), Barcelona, Spain}
\address[UIC]{Universitat Internacional de Catalunya (UIC), Barcelona, Spain}
\address[UDD]{Data Science Institute, Universidad del Desarrollo, Santiago, Chile}
\address[IAAC]{Institute of Advanced Architecture of Catalonia (IAAC), Barcelona, Spain}

\begin{abstract} 
Life, services and activities within cities have \colorA{commonly} been studied by separate disciplines, each one independent from the others. One such approach is the computer simulation, which enables in-depth modelling and cost-effective evaluation of city phenomena. However, the adoption of integrated \colorA{city} simulations faces several barriers, such as managerial, social, and technical, despite its potential to support city planning and policymaking. 

This paper introduces the City Physiology: a new \colorA{conceptual} framework to facilitate the integration of city layers when designing holistic simulators. The physiology is introduced and applied through a process of three steps. Firstly, a literature review is offered in order to study the terminology and the progress already made towards integrated modelling of different urban systems.
\colorA{Secondly, interactions between urban systems are extracted from the approaches studied before. Finally, the pipeline to carry out the integration strategy is described.}

In addition to providing a conceptual tool for holistic simulations, the framework enables the discovery of new research lines generated by previously unseen connections between city layers. Being an open framework, available to all researchers to use and broaden, the authors of this paper envisage that it will be a valuable resource in establishing an exact science of cities. 

\end{abstract}


\end{frontmatter}


\section{Introduction}
\label{sec:Introduction}

According to the \citet{populationdivisionoftheunitednationsdepartmentofeconomicandsocialaffairsundesa2018RevisionWorld2018} of the United Nations, in 2018 55\% of the world population was living in an urban environment, demonstrating a significant growth in the urban population from 30\% in 1950 to a forecasted 68\% by 2050.
The  growth of urban populations predicted in the next three decades leads to an urgent need to focus on the evolution of the urban environment, in order to assist researchers and policymakers in the definition of the future city. How cities will grow, whether in extension or density, is now uncertain, but understanding this process is critical to define future cities as more livable places.

Virtual cities are useful tools for both the management and administration of cities, and for the study of socio-economic phenomena.
Moreover, simulations can support planners in the evaluation of both existing and future city policies.
Computer simulations are the test field where planners can evaluate the behaviour of urban systems when exposed to changing conditions: in fact, the impacts of changes in city policies can be observed, alternative planning strategies can be explored, and new development tactics can be studied \citep{kilbridgeConceptualFrameworkUrban1969}. 
This process includes, but is not limited to, the tasks of: the assesment of both foreseen and unforeseen impacts, consequent conflict prediction and/or understanding, and the optimisation of policies \citep{coutureIncomeGrowthDistributional}.
The circular relationship between theory and modelling \citep{kilbridgeConceptualFrameworkUrban1969} is particularly relevant in studying how a city functions. In the test field, the researcher can experiment with different relationships between components 
and discover which parameters are more relevant in explaining reality. Thus, simulations also help in defining the theory behind the studied phenomenon. This test field allows the investigation of scenarios --- cost-effective and in-depth --- that would be costly, unethical, or simply impossible to carry out in reality. Hence, simulations can serve as a tool to better understand urban dynamics --- a highly complex system with feedback mechanisms which are difficult to predict.

The traditional approach to simulations of urban dynamics is to resolve issues only for an individual city layer or, at most, for partially coupled layers. Integrated urban models have been adopted to investigate land use and mobility, but whether to adopt them in real urban environments still depends on their usefulness in the planning activity \citep{millerIntegratedUrbanModeling2018,Graells_Garrido_2020}. Even though specific sectors are increasingly interested in enriching their simulations, specifically by integrating layers, they lack a holistic approach.
Indeed, there is undeniable interest in applying complex system theory and network science to city problems~\citep{wallothUnderstandingComplexUrban2014}. Nevertheless, it can be said that there is no cultural-social agreement on the way of applying complexity studies to cities. This paper aims to contribute to the complex definition of cities.
\colorA{City digital twins, such as the one of Herrenberg \citep{dembskiUrbanDigitalTwins2020}, share the same purpose of this article, but with a different approach: the digital twin infrastructure consider layers in an pragmatic manner, when data and simulation models become available for technical connection and implementation. Instead of following a conceptually-solid integration, layers are connected ad-hoc, therefore the holistic approach of the city digital twin is valid locally, but not universally.
On the other hand, they prioritize features such as real-time data streaming and powerful visualization.}

Why are holistic simulations of a city not common? Is it considered to be only a technological challenge? Is it worth the effort? It has been acknowledged that, even when integration is key to improve simulations, there are management, social, and technical barriers \citep{bachCriticalReviewIntegrated2014} that restrain the adoption of both integrated models and models per se (even if not integrated). 
These barriers can be classified into four groups:
model complexity, user friendliness, administrative fragmentation, and communication \citep[Table 4, page 102]{bachCriticalReviewIntegrated2014} .

What seems to be evident is that different fields of urban studies adopt simulation techniques, independently of one another, yet they try to include input from other disciplines. This approach leads to having to deal with domain-knowledge silos and technical limitations. Since a city is where countless life activities are condensed, different aspects of a city relate to numerous disciplines, without explicitly acknowledging their belonging to the well-being of the city \citep{battyConfusionTerminologies2019}.
Moreover, a globally shared idea of city is needed: a collection of precisely named components to be modelled. Not only does its complexity engender multiple interpretations of the idea of city, which then loses its universality, but the description of its components may also be incomplete: in fact, the relationships between components are not always clearly defined, and how they affect each other can be unclear.

We believe that the city should be represented by a holistic approach: not a simple sum of the parts, but a system the properties of which emerges from the effects of each layer on all the others.
\colorA{The City Protocol Society  (\citeyear{cityprotocolsocietyCityAnatomyFramework2015}) share the same 
holistic view, and serves as reference for defining city layers.}
\colorA{Note that the terms system and layer are used indistinctively in this paper, due to the several usages of these concepts in the literature.} 

The contribution of this paper is a framework that allows: 
\begin{itemize}
    \item the review and comparison of simulations from different research fields, that can be organized according to their contribution to urban systems (as defined by the City Protocol Society);
    \item the provision of a qualitative review according to the field in question (`\nameref{city-sys-and-simu}' section);
    \item the direct spotting of which are the relevant interactions between urban systems for improving holistically the simulation, as much as new research lines opened by missing connections (`\nameref{developing-int-multisim}' section).
\end{itemize}
Firstly, each urban system is studied individually. Secondly, how the numerous different sectors show interest in integration is described (e.g. water modelling, urban metabolism, air quality) although, up to now, no example of holistic simulation, \colorA{matching the authors' standards of replicability and conceptually-solid integration,} is found to be in use. 
\colorA{Then, interactions among urban sub-systems are extracted from urban models.} Lastly, remarks and ideas for further development of this line of research are offered.

\section{City systems and simulations}
\label{city-sys-and-simu}

\colorA{In this section the definition of city is explained. The state of simulation models for each urban system is reviewed accordingly.
For each urban system, references were found following a search criterion that is a combination of: the layer of interest (e.g. energy, water, communication) + integrated + urban + simulation/model/modelling. Only review papers from the last 5 years were considered (when possible), together with other publications found during the development of this work and that the authors considered relevant.}

\subsection{City definition}\label{city-definition}

Given its complexity, there is no commonly recognized definition of \textit{city}; in fact the concept is shaped around political, economic and social ideas \citep{cityprotocolsocietyCityAnatomyFramework2015}, as much as around urbanistic and architectural ones. Our reference definition is given by the `City Anatomy' \citep{guallartSelfSufficientCityInternet2014, cityprotocolsocietyCityAnatomyFramework2015}: the city is `a system of systems and interactions that fosters emergent human behavior'. This system is based on the relationship between multiple layers of a settlement, that are administered and regulated according to a system of laws, both local and recognized worldwide. 

In the `City Anatomy' the City Protocol Society (CPS) (\citeyear{cityprotocolsocietyCityAnatomyFramework2015}) makes an analogy between the city and the body: it introduces the holistic rule that the ensemble of the parts is more that their sum, with complexity arising not from the sum of all the systems, but from their capability to function in coordination.

The systems considered by the CPS (\citeyear{cityprotocolsocietyCityAnatomyFramework2015}) as basic and common to every city are six infrastructure layers that enable flows from/to and within a city (\textit{Communication Network, Water Cycle, Energy Cycle, Matter Cycle, Mobility Network} and \textit{Nature}), the \textit{Built Domain}, \textit{Society}, and \textit{Environment} (see \nameref{SI_Appendix_anatomy} for further details.) However, the CPS (\citeyear{cityprotocolsocietyCityAnatomyFramework2015}) only explains the theory framework and the system subdivision, and not how each system (or layer) interacts with the others, even though they define indicators (\citeyear{cityprotocolsocietyCityAnatomyIndicators2015}) for each layer and a high level ontology (\citeyear{cityprotocolsocietyCityAnatomyOntology2015}).

The remaining steps are not simple: indeed, just defining the elements and interactions of a holistic city simulator is a challenge by itself. This defining process comprises three steps: naming, describing, and modelling. 
The first step consists in recognizing and naming firstly each system, and then the elements of each system. Collecting and naming all the elements that should be considered requires in-depth investigation, but it is not sufficient to enable success in the modelling phase, that focuses on how to make the interactions work (interactions that need to be named as well). The second step consists in describing how the systems work and interact theoretically, that is to say, the interaction of elements within and between systems. This is the phase in which this article is. Returning to the metaphor of the city as a body, while the anatomy focuses on the structure and components of the body, this paper proposes to develop a \textit{City
Physiology}: how these components work and relate to each other.

The City Anatomy \citep{cityprotocolsocietyCityAnatomyFramework2015} is a `framework to support City Governance, Evaluation and Transformation'. 
It seeks to define `a common systems view for all cities regardless of size or type' and to interconnect them through an `internet of cities'. The choice of this particular framework as a definition of the city for building a holistic simulator comes from its ISO certification \citep{isoISO3710520192019} and its being a relatively recent publication (less that 10 years), and being supported by one of the authors of this paper.

Finally, the division of the problem into a system of coupled systems is advantageous from a computational perspective, where a divide and conquer approach with parallel algorithms communicating at their boundaries is a natural solution for today's high performance computing environments.

\subsection{Urban modelling}\label{urban-modelling}

\colorA{On the confusion of the terminologies \textit{modelling/model}, whose meaning vary among urban-related disciplines, the \nameref{SI_Appendix_models} defines a common glossary. 
In this paper a model is understood to be a \textit{simulation model}. The model is the abstraction that a simulator uses to run the simulation, so the \textit{simulator} is the engine that runs on the (simulation) model \citep{yilmaz2004conceptual}. Both simulation models and simulators were considered for reference, and the search criteria (`simulation/model/modelling') explained at the beginning of the section take this into account.
}

The \textit{urban modelling} literature focuses on specific elements of the city: models of mobility, land use and urban growth.
Ubication theories \citep{battyUrbanModeling2009} (where and why economic activity takes place) together with physical space, land use, population, employment, and transportation concepts are commons to many definitions of urban modelling and models. (See \nameref{SI_Table_modeldefinition} for definitions).
Worthy of note are: `urban planning models' by \cite{kilbridgeConceptualFrameworkUrban1969}, `LUTI -- land use transportation interaction' models \citep{battyGenericFrameworkComputational2012}, and the more recent definitions of `large-scale urban models' \citep{leeRetrospectiveLargeScaleUrban1994}, Batty's `urban models' (\citeyear{battyUrbanModeling2009}) and `IUMs -- integrated urban models' by \cite{millerIntegratedUrbanModeling2018}. In addition, the urban phenomena should be described in structural and mathematical terms \citep{kilbridgeConceptualFrameworkUrban1969, battyUrbanModeling2009, leeRetrospectiveLargeScaleUrban1994}. With his definition of `Urban Dynamics' \cite{battyUrbanModeling2009} generalizes on the `myriad of processes at work in cities' through time, but always relative to the changes in the spatial structure of the city. 

The city as an entity, its services and its activities, has always been the subject of study by different disciplines, but only recently different fields of the academia explicitly connect these interests with a single discipline studying the city \citep{battyConfusionTerminologies2019}; think of terms like \textit{urban analytics}, \textit{city science}, \textit{digital twin} or \textit{\colorA{geographic information system} (GIS)}.

\subsection{Urban form, land use and mobility
}\label{urban-form-land-use-and-mobility}

The \textit{urban modelling} concept starts from a precise problem --- traffic --- and gradually includes demographics, employment and urban form. Nevertheless, it does not achieve an inclusive view of all the elements of a city.
Traffic and expensive transport infrastructures cost-benefit analysis is the problem that awoke interest in \textit{urban simulations} in the 60's and 70's \citep{millerIntegratedUrbanModeling2018}. Mobility simulations were enriched with land use information: first, demographic and employment data fed the mobility model; next, the impact of the new infrastructure on land values was studied, then, how the changed values affected the infrastructure was analyzed. 
See \cite[Figure 1 - The land-use transport feedback cycle]{wegenerLandUseTransportInteraction2004} for a clear representation of mentioned interaction.

\citet{liUrbanGrowthModels2016} consider `urban growth' as involving `economics, geography, sociology, and statistics' and a variety of activities such as `land use, housing, population, travel, networks, transport, employment, workplaces'. Two lines of evolution are followed: from `macro' to `micro' (spatial dimension) and from `static' to `dynamic' (temporal dimension). In their review they define three groups of models: 
\begin{itemize}
    \item LUT models (focusing on socio-economic activities);
    \item \textit{cellular automata} (CA)-based models (for land use);
    \item \textit{agent-based models} (ABM) (for both).
\end{itemize} 

Early predictive models are `static' models because they seek equilibrium for a specific point in time \citep{berling-wolffModelingUrbanLandscape2004}. They are based on the theories \citep{battySciencePlanningTheory2017} of social physics, geographic distribution (of demographic groups, density and work activity), and economic utility. 
These models run macro-analysis \citep{kilbridgeConceptualFrameworkUrban1969} that study mobility on aggregated level (gravitational theories) and how this mobility affects land use (geographic distribution), instead of trips through a network \citep{berling-wolffModelingUrbanLandscape2004}.
Micro-analysis decision models study the rational choice of the individual \citep{kilbridgeConceptualFrameworkUrban1969}, who chooses between a discrete number of alternatives \citep{berling-wolffModelingUrbanLandscape2004}. Alternatives are formulated on market and location theories (economic utility); consequently, trips and traffic depend on dwelling decisions that are a trade-off between distance from workplace and quality of space.

Since cities are rarely in a state of equilibrium \citep{berling-wolffModelingUrbanLandscape2004}, models with dynamic characteristics are needed and have been developed; these models consider the temporal aspect of the studied phenomena, so they have a memory of past states of the system.
CA are among the first models developed: they consist of systems of cells where the state of each one changes according to the state of adjoining cells \citep{liUrbanGrowthModels2016, berling-wolffModelingUrbanLandscape2004}. 
The interaction is simple, but, it causes local decisions to spread across the entire system and to display large-scale patrons together \citep{berling-wolffModelingUrbanLandscape2004}. 
The structure of CA leads to the development of \textit{Spatial analysis} techniques which study: variables that change significantly through space, spatial correlation between data, and how the introduction of additional space changes the hypothesis of a model \citep{berling-wolffModelingUrbanLandscape2004}. CA can be considered as a simplification of ABM: both are disaggregated models, representing the behaviour of individual located in space. 
ABM generally works on a micro-scale and on specific questions, e.g. pedestrian or vehicular traffic, housing market policy, and segregation \citep{battyUrbanModeling2009}.

\citet{berling-wolffModelingUrbanLandscape2004} also mention two other theories: fractal forms (that has its own niche) where the growth dynamic depends on the geometry of the the system, and \textit{ecological energetics} theories (urban metabolism models) that consider the form of the city dependent on exogenous energetic inputs and on its internal capability to renewal.

Since the 90's, neural network theories have been incorporated within urban models (not sufficiently described by dynamic systems): instead of adapting urban theories to mathematical modelling, they attempt to adapt mathematical methods to urban processes too complex to be described with a system of equations \citep{berling-wolffModelingUrbanLandscape2004}.

Urban form, land use, and mobility are deeply connected. In the subsection above different spatial analysis techniques (e.g. CA, ABM) are mentioned and how they evolved both on the spatial dimension (macro/micro) and on the temporal one (static/dynamic) is explained.

\subsubsection{Built domain }\label{built-domain}

The Built Domain has many scales, and the technology that best represents the holistic approach concept at the building scale are \textit{Building Information Models} (BIM). Instead of specialising, BIMs seek a well-rounded and integrated vision of constructions (geometry, materials, costs, time, management and maintenance).

\citet{wangIntegrationBIMGIS2019} posit that BIM `focuses on micro-level representation' but can be coupled with GIS, that provide `macro-level representation of the external environments of buildings'. Their review focuses on technical integration (IFC and CityGML semantic models, information conversion and loss) and shows different applications in `energy management', `urban governance', `life cycle of architecture, engineering and construction projects', and `data integration'. Thanks to GIS integration, BIM models have an urban-scaled context and can be integrated with other systems (such as the Energy Cycle).

`Computer-based form generation' techniques at a building scale are reviewed in \citep{grobmanComputerBasedFormGeneration2009}. The authors investigate methods, developed both in research and practice, that can be summed up as: activities allocation and complex formal patterns problems (e.g. CA, as mentioned for land use problems as well); composition rules-based methods, based on shape transformation (e.g. `Shape grammar') or performance data.

Both BIM and GIS technologies allow, at a different scale, an integrated vision of city elements.
The review analysis led to modelling techniques already encountered for other layers. Nonetheless, in this case they're applied at a different scale because of the nature of this layer, that consider elements usually not at city scale.

\subsubsection{ Mobility network }\label{mobility-network}

\citet{kiiTransportationSpatialDevelopment2016} review LUTI models, according to spatial dimention:
local-scale intensive modelling and regional --- or global --- scalesimplified modelling. Like other authors, they divide LUTI models into three categories: 
\begin{itemize}
    \item `Spatial interaction/gravity-type modelling';
    \item `Econometric models' (the spatial allocation of activities follows random utility theory, rather than a gravity model);
    \item `Microsimulation
and the other computer-based models' (activity-based highly
disaggregated approach).
\end{itemize}

Among the `other computer-based models', \citet{kiiTransportationSpatialDevelopment2016} describe the `incorporation of social interaction modelling', where behavioral choices are seen through the inclusion of social network structure, and information and communication technology (ICT).

Following this perspective, many studies --- done using the growing amount of data coming from the Communication Network --- lie at the borders between mobility, social and computer science, and ICT. 
Some examples worthy of mention: people mobility \citep{graells2018inferring} and segregation \citep{beiroShoppingMallAttraction2018} are inferred from mobile phone location data, and georeferenced Twitter data is analyzed for land use \citep{frias-martinezEstimationUrbanCommuting2012}, touristic attractiveness \citep{bassolasTouristicSiteAttractiveness2016}, and human perception of a place \citep{crooksCrowdsourcingUrbanForm2015}.
In contrast with previous city layers,
the examples found here are data-driven works,
which could be used to define more realistic models in simulations.

\subsection{Communication network }\label{communication-network}

The search for simulation models of the Communication Network led to technical surveys, which are focused mostly on the performance of wirelessnetworks. 
The review papers which were found lack a city-wide approach, but they are widely used to validate wireless network before real-world deployment
\citep{khanPerformanceComparisonOpen2012, minakovComparativeStudyRecent2016}.

However, in considering the data generated by the Communication Network, its interaction with other systems of the city is essential, especially in the study of human behaviour within the urban context.
\citet{ilievaSocialmediaDataUrban2018} conducted a deep review on the use of `spatial and semantic information from social media data' in the research and planning of urban-sustainability. They identify five main topics: environmental sustainability, public health, social equity, mobility and economic development.

As a future work, it would be interesting to extend the literature research to understand how other layers of the city can impact on the network functioning (e.g. how the Built Domain affect wireless communications.)

\subsection{Society }\label{society}

According to \citet{vanbavelIntroductionAgentBasedModelling2017}, demographic studies rely on two concepts of population:
the `national population' (the bookkeeping of phenomena as fertility, morbidity, mortality, birth-rate, and migrations) overtook the `open' approach to population, that focuses on the `processes and mechanisms that generate patterns of association between individuals, such as mating or social networks, and how these processes affect population change and heterogeneity'. 
Consequently, the authors arrived at the conclusion that using ABM is useful to connect the two concepts.

According to \citet{courgeauModelBasedDemographyResearch2017}, ABM falls within the `social simulation' trend of demography, which focuses on applying `novel modelling techniques to specific populations or simulations'. Numerous ABM are bottom-up models, where some low-level interactions supposedly produce high-level complex behavior. However, the authors underline the necessity of multilevel analysis (micro and macro at the same time) because of the emerging properties of complex systems, where the aggregation of micro-level patterns cannot generate macro-level patterns. They state that population science should focus on the interaction between different population systems (individuals, groups, and institutions) and on how the multilevel interactions shape demographic outcomes. However, the authors recognize the shortcomings of model-based approaches.
See the book `Simulation for the social scientist' \citep{gilbertSimulationSocialScientist2005} for the common approaches to social simulations.

The Society system shows a growing cohesion with all the other systems. Indeed, city dwellers are ultimately the end-users: either for their demand (land use, transport, water, and other resources) or for the impact they suffer (water, air, noise pollution).
In fact, the idea of activity-based models (the activity of the individual is the unit of analysis) is well established within the LUT models and it is borrowed by other systems too --- like the Energy Cycle --- \citep{keirsteadUsingActivityBasedModeling2012}.
Therefore, even for analysis typically done at the scale of the city ---
like urban metabolism --- including the resource demand at a finer
scale can help in design better infrastructure and understand the impact
of policy interventions.

\subsection{Water, energy and matter}\label{water-energy-and-matter}

\citet{hussienIntegratedModelEvaluate2017} explain that disaggregating the water-energy-food (WEF) nexus into end-use consumption (household scale), is the most appropriate scale to identify best and worst practice, and consequently influence consumption behavior.

\citet{newell40yearReviewFood2019} analyze different papers focusing on the WEF nexus, also investigating whether multiscale interaction and political and socio-economic factors affecting the flows are considered or not, in particular the role of institutions. 
The second half of their paper is dedicated to WEF systems at the urban scale, using the same approach as above. They propose using the `urban metabolism concept as the interdisciplinary frame to integrate interactions, disciplines and stakeholders'.

Urban metabolism studies are very close to the idea of integration of city elements: metabolism is the sum of all the technical and socio-economic processes within a city, that determine growth, energy production and waste disposal (see \textit{Material flow analysis} and Odum's \textit{Emergy} theory) (Kennedy in \citet{kennedyStudyUrbanMetabolism2011}).

Marx introduces the concept of urban metabolism (though politically). However Wolman is the first to attempt to quantify it, by focusing on the human demand for resources, with his hypothetical city of one million people \citep{kennedyStudyUrbanMetabolism2011, chrysoulakisSustainableUrbanMetabolism2013, pincetlExpandedUrbanMetabolism2012}.

Real-application studies on urban metabolism began in the 70's and they followed two non-conflictual schools: the \textit{Material Flow approach} and Odum's energy-equivalent \textit{Emergy}. The flow/mass balance method is more widespread because it is more flexible and allows to use different units. In the 90's, Newman connected urban metabolism measures with livability and sustainability analysis, expanding urban metabolism to urban system analysis \citep{kennedyStudyUrbanMetabolism2011, pincetlExpandedUrbanMetabolism2012}.

Urban metabolism research uses the double analogy of a city as an organism and as an ecosystem \citep{kennedyStudyUrbanMetabolism2011, chrysoulakisSustainableUrbanMetabolism2013, pincetlExpandedUrbanMetabolism2012}: the organism consumes resources from the environment and returns waste, with a linear one-direction metabolism of material and energy fluxes; the ecosystem hosts different organisms (people, animals, vegetation), and natural ecosystems are generally auto sufficient (like the appealing idea of a sustainable city). 

Typically used for energy and material balance, this method supports the urban planning activity, because it can evaluate environmental problems and incorporate socioeconomic and policy analysis \citep{chrysoulakisSustainableUrbanMetabolism2013,pincetlExpandedUrbanMetabolism2012}. 
In fact, according to \citet{pincetlExpandedUrbanMetabolism2012}, the political and economic forces that affect how cities grow are intrinsic to the urban metabolism, which is complex and integrated with both local and global systems.

Therefore, sustainable urban planning requires a systems approach which is able to fill communication gaps. Depending on the scope of the study, different aspects of the same variable are relevant (e.g. energy: policy-makers focus on consumption; meteorologists focus on how it is stored and transmitted to influence microclimate data). 
A system-wide approach is necessary to match variables from different systems: integration is difficult because it requires the harmonization of units, scale and boundaries, and then their correlation to human decision-making. Numerous studies use highly aggregated data at a regional level that cannot represent local peculiarities (top-down approach). Furthermore, this approach also depends on data availability: data integration should be a common commitment by political, economic, and institutional entities \citep{chrysoulakisSustainableUrbanMetabolism2013, pincetlExpandedUrbanMetabolism2012}.

Water, energy, and matter are deeply connected; this nexus is studied in depth in the literature reviewed. Urban metabolism studies present a good level of integration between city subsystems and its two core theories are listed above in the subsection.

\subsubsection{Water cycle}\label{water-cycle}

The European Water Framework Directive (2000) is cited by many authors as a critical point in legal direction and also represents a shift in the approach to integrated modelling approach. 
In the water modelling field, a switch occurred in the boundaries definition during the 1990s and 2000s.
`Coherent boundaries' were replaced to adapt to a new concept, as later translated into the new legislation introduced by the European Directive.
The switch is from a boundary based `emission point of view', such as for treatment plants and sewer systems, to a new one where water quality outcomes set the boundaries. 
Nowdays, the focus is set by the EU Framework, and it aims to reach `good ecological status' for surface water, emission control, and environmental standards. 
According to the emission source paradigm, point sources are single points (such as a factory or a sewage treatment plant), while diffuse sources are precipitation, atmospheric deposition, drainage and water from land runoff, and even detritus from urban areas \citep{bachCriticalReviewIntegrated2014, biegelArcEGMOURBANHydrologicalModel2005, vanrolleghemContinuitybasedInterfacingModels2005}.

As previously seen for different aspects of the city, a problem of scale occurs: river basins do not follow administrative boundaries. 
Furthermore, integration adds to the scale issue: it becomes necessary to model individual catchment processes as much as their integration, a context in which frameworks such as the OpenMI \citep{gregersenOpenMIOpenModelling2007} are developed.
However integration has to deal with these often unsolved questions:
missing variables, their meaning, and the elemental composition (the chemistry) \citep{vanrolleghemContinuitybasedInterfacingModels2005}.

A quest for  optimization is needed: integration is required in order to achieve new water management goals such as water recycling and the interaction with urban space (micro-climate control, amenities protection, flooding control). The new requirements encroach on social and economic domains, reaching the highest level of integration as defined by \colorB{\citet{bachCriticalReviewIntegrated2014}} in their classification.

As mentioned before, the INTERURBA I conference focuses on drainage systems and integration, but all the attention is always on the drainage: even if the water system belongs to a wider frame (such as river basin and ecological systems) the literature about integration historically focuses on `urban water' systems drainage.
However, the authors acknowledge that many integration concepts developed are suitable to be sufficiently generalized to use with other aspects of the water system.
The water supply system is less studied because it's a more controllable system, with probably a satisfactory level of performance and only slightly affected by natural phenomena \colorB{\citep{bachCriticalReviewIntegrated2014}}.

\citet{fuReviewCatchmentscaleWater2019} review catchment-scale water quality models of non-urban systems, with `emphasis on sediments and nutrients',
while \citet{luReviewSocialWater2016} review the Social Water Cycle (SWC) field, that 
consider the impact of human activities as a component of the natural water cycle. Therefore, by comparing previous and modern water systems, they believe studies have gradually shifted their focus from resource allocation to limiting the consumption. Again, the management scope has been broadened from supply and drainage to `including river health, transportation, entertainment facilities, micro climates, energy sources, grain yields, etc'. They conclude that, even if the socio-economic system has a great impact, it could still risks being seen as a collateral factor, so they envision future work on the understanding of fundamentals.

\citet{desouzagroppoPredictingWaterDemand2019} describe the challenge, that big cities have recorded in particular, to provide sufficient potable water, and review urban water demand forecasting methods that employ artificial intelligence.

Water models include drainage and supply systems. As explained above, they can be classified according to the level of integration of their subsystems and whether or not they have an interdisciplinary angle. Socio-economic variables have indeed a significant impact on water systems. The forecasting methods reviewed bring the attention on the water availability issue.

\subsubsection{ Energy cycle }\label{energy-cycle}

In the review of energy systems by \citet{subramanianModelingSimulationEnergy2018} we find, again, examples of ABM and neural networks modelling approaches, here used to model supply chain entities, electricity markets and technological change.

\citet{wangIntegrationBIMGIS2019} review energy infrastructure models (electric
power, natural gas, and fuel networks) from the resilience point of view
(resilience indicators).

\citet{abbasabadiUrbanEnergyUse2019} argue that the tools for urban energy modelling often focus on singular components and lack an integrated approach. 
They focus their review on urban energy, in the form of:
`building operational energy', `building embodied energy', `transportation energy', and `road and infrastructure energy' (this including also the life cycle effects of material flows). They divide estimation methods into top-down (macro scale) and bottom-up models (individual units), and focus on the latter, subdivided in:
`data-driven' (`statistical methods', `artificial intelligence-baes models') and `simulation' methods. They conclude by proposing a hybrid data-driven and simulation approach.

\subsubsection{ Matter cycle}\label{matter-cycle}

Previously, how material flows affect infrastructure energy emerged (in the form of energy for transportation): the Matter Cycle is indeed also a matter of transports. The CPS (\citeyear{cityprotocolsocietyCityAnatomyFramework2015}) argues that food is of utmost importance in the Matter Cycle because almost a third of the production is lost in waste. Therefore waste management become pivotal in this research, which studies the optimization of the waste collection routing problem.

\citet{belienMunicipalSolidWaste2014} present a review on solid waste management and vehicle routing problems: papers are classified into different categories with the aim of facilitating further research. The categories range from type and disposal facilities to modelling solution and types of constraint being added to the model.

To conclude, from WEF nexus reviews, the impact from/on food systems related to water and energy systems is understood.

\subsection{Environment }\label{environment}

Given that the CPS (\citeyear{cityprotocolsocietyCityAnatomyFramework2015}) considers the environment as the physical setting where cities are established, this research studies the field of eco --- from `oikos' = home --- systems.

The already cited H.T. Odum (father of the Material Flow approach for urban metabolism) was a system ecologist. In fact \citet{weilandUrbanEcologyBrief2011} cite the `analyses of urban material and energy flows' as one of the two main streams developed within ecosystem studies within `urban ecology'.

The field of `urban ecology', as explained in \citep{weilandUrbanEcologyBrief2011}, is an `interdisciplinary research field at the interface of natural sciences, social sciences and humanities as well as engineering'. Indeed, it `investigates the interrelations between environmental compartments and human activities such as construction, production, housing, and transport'.

Thus, it can be said that the `City Anatomy' \citep{cityprotocolsocietyCityAnatomyFramework2015} shares the scope of urban ecology: studying the urban environment, its components, and the activities that take place there.

\subsubsection{Air}\label{air}

In the vision of the CPS (\citeyear{cityprotocolsocietyCityAnatomyFramework2015}), the Environment includes: nature (outside the city borders) and air, soil and water. Nature, soil and water have their urban counterparts within the `infrastructure layers'; for this reason, this research focuses on air simulation.
Air is critically important for quality of life - indeed, air quality belongs to the indicators of both the 3rd and the 7th Sustainable Development Goals shared by the United Nations \citep{inter-agencyandexpertgrouponsdgindicatorsGlobalIndicatorFramework2017}) - and it is a well-established field of research in the computational simulation domain.

As \citet{marquezFrameworkLinkingUrban1999} point out, the structure of several cities encourage different unsustainable practices for resources (air, water, energy) and waste production. For instance, the zoning practice creates a dependency on road-based traffic. Indeed, air quality studies focus on (i) reducing pollutant concentration (to optimize emissions) and (ii) optimizing exposure (which areas of the city people should live in). Consequently, the authors question whether the compact city is the best form to improve the quality of life within cities. Therefore, it becomes mandatory to consider air quality and other environmental indicators alongside the process that, traditionally, is seen as a strictly land use and transportation problem. (Other examples of improving urban growth problems with environmental indicators are cited in the `\nameref{nature-infrastructure}'  section).

WHO guidelines on pollutants limits (e.g. NOx, Ozone or PM10) are not met in numerous cities, even when exposure to pollution has a high cost not only in health terms, but also in agriculture, natural ecosystems, the deterioration of building materials, and other aesthetic aspects of the environment \citep{benavidesCALIOPEUrbanV1Coupling2019, gurramAgentbasedModelingEstimate2019, marquezFrameworkLinkingUrban1999}. The impacts of pollution are not equally distributed, affecting socially and economically disadvantaged people differently.

How urban form affects the sustainable development of cities in the interaction between policy making, air pollution, and contaminant exposure \citep{marquezFrameworkLinkingUrban1999, gurramAgentbasedModelingEstimate2019} must be considered. 
Therefore, data at street level is needed to develop concrete actions, e.g. to choose walking paths through less polluted streets or to reduce traffic near sensitive locations such as hospitals and schools. The level of detail required for such insights must overcome different barriers: 
(i) the cost of sensors technology,
(ii) the poor density of this monitoring stations, or alternatively
(iii) the variable quality of cheap sensors, and 
(iv) the high resolution (1km x 1km) of mesoscale air quality modelling \citep{benavidesCALIOPEUrbanV1Coupling2019, gurramAgentbasedModelingEstimate2019}.
\colorA{The digital twin of Herrenberg \citep{dembskiUrbanDigitalTwins2020} combined air flow and mobility simulations with the aim of concrete actions in mind, to support the sustainable governance of the city.}

Mesoscale pollution models like HERMES attempt to include as many different sources of pollution as possible \citep{benavidesCALIOPEUrbanV1Coupling2019}. In fact, better air
quality initiatives are connected especially with traffic, as well as to other hot topics of increasing public concern such as harbour activity --- see \citep{tichavskaPortcityExhaustEmission2015} for specific analyses.

To understand the conceptual and technical aspects related to urban air quality modelling, such as urban microclimate, thermal properties of urban canopy and pollutant emission, dispersion and transport, see \citep{kadaveruguHighResolutionUrban2019}.

Similarly to what has been seen for the Water Cycle previously, air quality simulations raise problems concerning scale and boundaries definition. 

\subsection{Nature infrastructure }\label{nature-infrastructure}

The use of flora in urban planning is common to respond to acoustic, air, water and soil pollution (noise barrier, gaseous absorption, phytodepuration and phytoremediation). Moreover, vegetation provides natural shading and temperature control within public spaces and for buildings (green canopy). Access to green spaces is important for recreational activities and psychophysical well-being (shareable gardens, biophilia). In hydrogeological risk management, vegetation reduces mountainside erosion.

As explained in the previous paragraph, the Nature Infrastructure has wide and diverse influence on multiple aspects of the city layers (Water and Energy Cycle, Society and Air) but it does not have autonomy within the layers of urban context, contrarily to the value it could have in rural areas.
This means that the simulation output does not belong to the Nature Infrastructure; even so, 
the peculiarity of this subsection as compared to the previous ones is recognised, 
and 
the branches of the simulation field that consider the Nature Infrastructure as the predominant input are studied.

Ecosystem services is one of them: \citet{linReviewUrbanForest2019} review urban forestry modelling case studies among different disciplines. To be considered in their review, these models had to use `forest structure' (size, species composition, spatial configuration) to estimate ecosystem services. The review focuses specifically on trees and shrubs. It excludes green roofs, green infrastructure and green space concepts, unless they focus on `the structure and benefits of urban trees and shrubs'. They analyse the case studies according to location, disciplinary field, modelling tools, and ecosystem services - that is economic, social or physical/biological benefits (such as removal of air pollutants, and temperature and microclimatic modifications).

Green park characteristics are the focus of the review by \citet{aramUrbanGreenSpace2019}, that studies the intensity of urban green space cooling effects and green space density. The main variables of influence are the size and the shape of the green parks, the vegetative cover, and the climate conditions, with large parks (more than 10 ha) giving the highest cooling effect distance and intensity. According to the authors, only a few studies consider the influence of non-vegetated surfaces within the park area (building, sidewalks) or the effect of natural and artificial elements (water bodies, furniture). Not many researchers take small parks into consideration, even though the studies that have focused on them have shown significant influence on the surroundings.

\subsection{Summary}
\label{summary}

\begin{figure}[!ht]
\includegraphics[width=\textwidth]{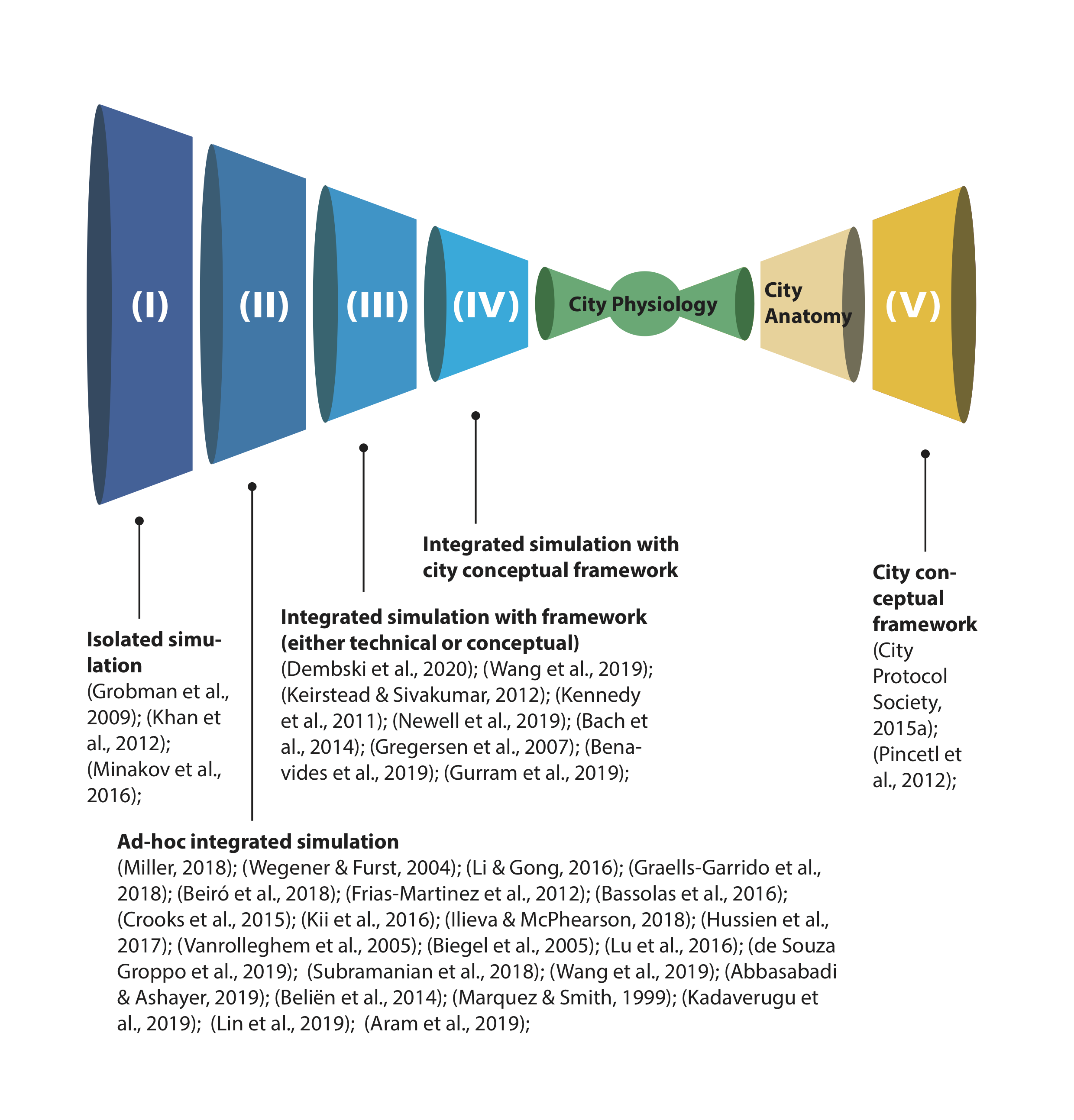}
\caption{\colorB{{\bf Towards holistic city simulations.}} \colorB{The references are categorized according to their degree of integration between layers. There are 5 level of integration: 
(i) Isolated simulation; (ii) Ad-hoc integrated simulation; (iii) Integrated simulation with framework, either technical or conceptual; (iv) Integrated simulation with city conceptual framework; (v) City conceptual framework.}}
\label{fig_funnel}
\end{figure}

\colorB{
The references are now categorized in Fig. \ref{fig_funnel} according to their degree of integration between layers. We identify 5 level of integration:
\begin{itemize}
   \item Isolated simulation: Problems are solved at layer level, not considering integration with other layers. Only a few references belong to this level, because a low-degree integration at least is usually considered.
    \item Ad-hoc integrated simulation: Different layers are coupled to solve the simulation with different degrees of integration, but the work is usually not general enough to be extensible to other layers. Most examples belong to this level.
    \item Integrated simulation with framework (either technical or conceptual): Coupled simulations within a framework. This is a step toward the holistic approach we are pursuing. However, this kind of research is usually focused on building a software or its architecture (technical framework), as in the case of the IoTwins digital twin \citep{borghesiIoTwinsDesignImplementation2021}. Alternatively, the coupling is based on sound theoretical concepts that, nevertheless, still lack some inputs to describe the city functioning holistically (conceptual framework), such in some urban metabolism studies.
   \item Integrated simulation with city conceptual framework: A city-wide conceptual framework is used to simulate the city as a whole. This problem is yet to be tackled but there’s a raising interest towards it, in fact the European Union is financing projects such as the H2020 IoTwins project \citep{DistributedDigitalTwins}. There’s a gap in the simulation of complex cities and, with this article, the authors want to start filling it.
    \item City conceptual framework: city-wide framework that consider the city holistically, such as the City Anatomy \citep{cityprotocolsocietyCityAnatomyFramework2015}.
\end{itemize}}

\clearpage

\colorA{Definitions of the concept of `city' and `urban modelling' are available in subsection \ref{city-definition} and \ref{urban-modelling} respectively. They are followed by a list, the most comprehensive possible, of examples of simulators, organized by the city layer they belong to. }

\colorA{
These examples are used to draw conclusions about interaction between city layers in the next section: `\nameref{developing-int-multisim}'.
Additionally, the typology of \nameref{SI_Appendix_models} is built upon the examples given in this section.}



\section{Developing integrated multilayer simulations}
\label{developing-int-multisim}

\colorA{The simulator conceived through the framework of this paper
} is a system of systems, both theoretically and computationally: in the first case, the structure and behaviour of the city are considered, in the second how the virtual city --- the simulator --- works.
\colorA{The behaviour and structure of the simulator are based on theoretical city models, e.g. the City Anatomy \citep{cityprotocolsocietyCityAnatomyFramework2015}, while the behaviour of every single system and the interactions occurring between them are conceived according to the theories developed by the sector experts, as explained in the `\nameref{city-sys-and-simu}' section.}

The aim of this vision is to convert the `\nameref{city-sys-and-simu}' collection of models --- made system by system --- into \textit{physiology} models. While the anatomy focuses on the analysis of the structure and the components of the city systems, the physiology studies how these systems work and are interrelated to each other.

The envisioned simulator is made up of systems and interactions:
\begin{itemize}
    \item \textit{System}: a system has a complex structure, but it is composed of elementary units which are based on simple enough rules that can be coded. These simple elements interact so that the complex behaviour can emerge;
    \item \textit{Interaction}: the interaction is a data exchange between different systems. It occurs between the input/output data of different systems. When data from System A work as input data for simulating System B, an interaction takes place. Furthermore, when at the same time, using data from System B improves the outputs of System A, then a strong interaction takes place, which is also called feedback cycle.
\end{itemize}

In the following \colorA{subsections}, input/output interactions between systems are described. 
The system-crossing interactions are extracted from the overview of the `\nameref{city-sys-and-simu}' section. 
The section ends with a consideration about data exchange \colorA{and the pipeline to apply the framework}.

\subsection{System-crossing interactions}\label{system-crossing-interactions}

The input/output interactions are explained following the city layers subdivision of the `\nameref{city-definition}' subsection:

\begin{itemize}
    \item Built Domain and Mobility Network systems interact through the well-known land use - transport feedback cycle. Additionally, land use data (in particular employment and housing location) can be used to model the Society system while soil properties affect the Water Cycle.
The designated use of buildings impacts on the Energy Cycle, while their construction, maintenance and demolition impacts on both the Energy and Matter cycles. The geometrical and thermal properties of the urban canopy affect air quality simulations and the cooling effects given by the natural surfaces;

    \item Mobility Network provides daily displacement and migration data that influence land use (Built Domain) and can be used to simulate the Society system. Transport of people and goods as well as the construction and maintenance of its infrastructure affect both Energy and Matter Cycle. Traffic emissions are environmentally critical for air quality simulations;

    \item Communication Network provides location data that are very relevant to study land use, mobility and segregation. Different metrics about Society can be extracted from social media information;

    \item the demographic data of the Society system is used to understand population mobility, both on a daily basis (displacement for work) and in the long term (household mobility and migrations). Households structures influence the demand and consumption of resources (Water, Energy and Matter cycles) and their behaviour shapes the response to policies managing the access to these resources. Specific sectors of the population (e.g. children, elderly, sick people and pregnant women) are considered to be sensitive population and they can influence air quality exposure and impact evaluations;

    \item apart from the WEF nexus, the characteristics of water bodies have a positive impact on air and cooling effect simulations. 
However, the concentration of pollution in water can have a significant negative impact on population health;

    \item Matter Cycle information on waste and residuals is useful both for water quality (pollutants) and mobility (collecting route) simulations;

    \item environmental information such as climate and weather data is significant for air and water simulations;

    \item information about air pollution concentration and distribution is useful data for modelling the Built Environment, the Mobility Network and the Society system. Air quality can shape urban growth by avoiding a morphology that worsens polluted air stagnation or that promotes car-dependent mobility behavior. In modelling mobility behavior, avoiding the more polluted areas or reducing traffic near sensitive locations can be considered;

    \item as for the Nature Infrastructure within the city, the green canopy affects microclimatic and air simulations.
\end{itemize}

\clearpage
\begin{figure}[!ht]
\centering
\includegraphics[height=0.9\textheight]{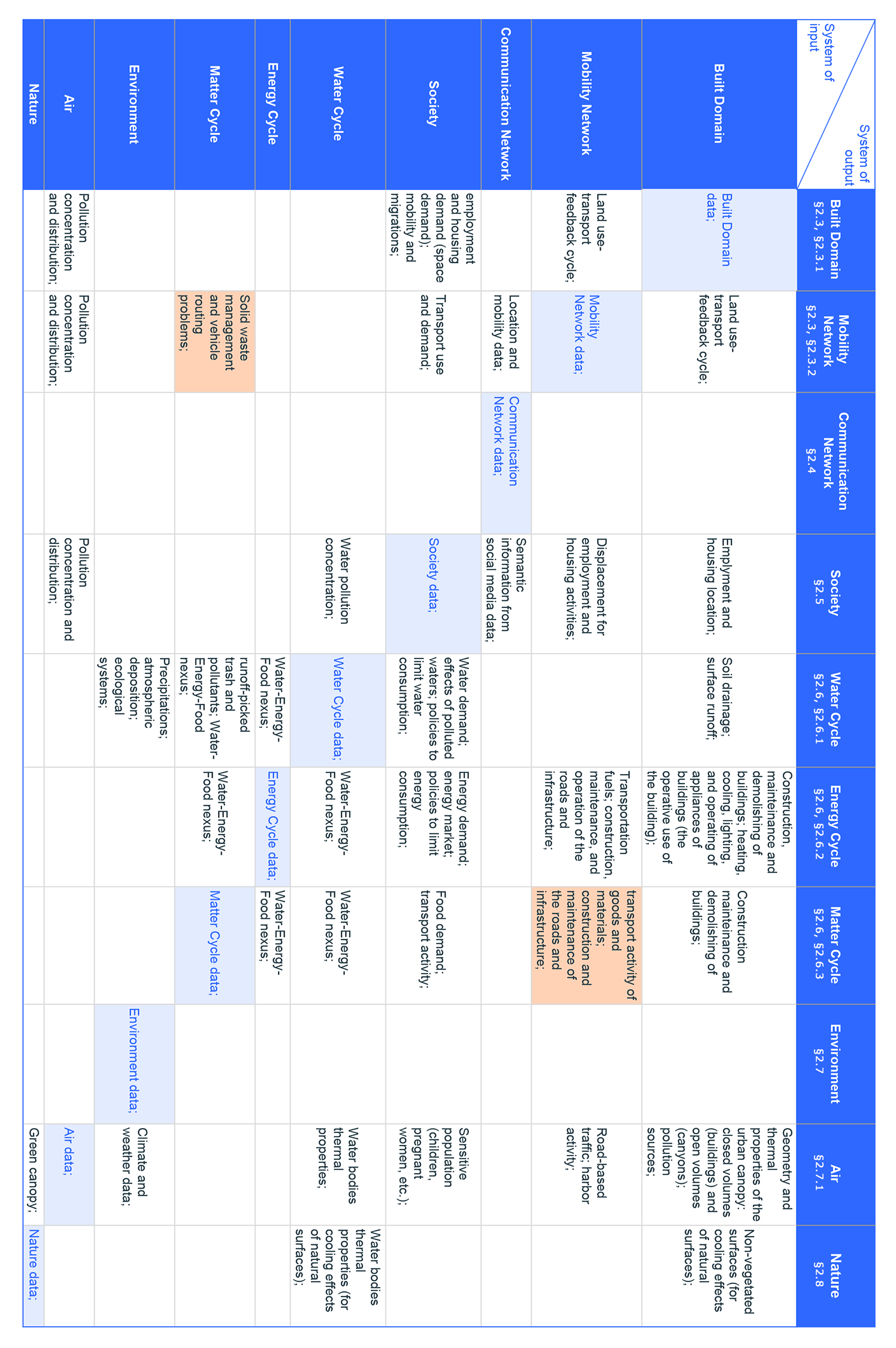}
\caption{{\bf System-crossing interactions and feedback cycles.}
In orange the cells taken as example in the paragraph
\colorA{ where the figure is first cited.}
\colorA{The input/output interactions are drawn from the literature analyzed in section \ref{city-sys-and-simu}, which is organized following the city layers subdivision. In the table each layer has the reference to the corresponding sub-sections.}
An editable version is available at \citep{irem_im_2020_4014187}.}
\label{fig_connections_table}
\end{figure}

\clearpage
\begin{figure}[!ht]
\centering
\includegraphics[width=0.8\textwidth]{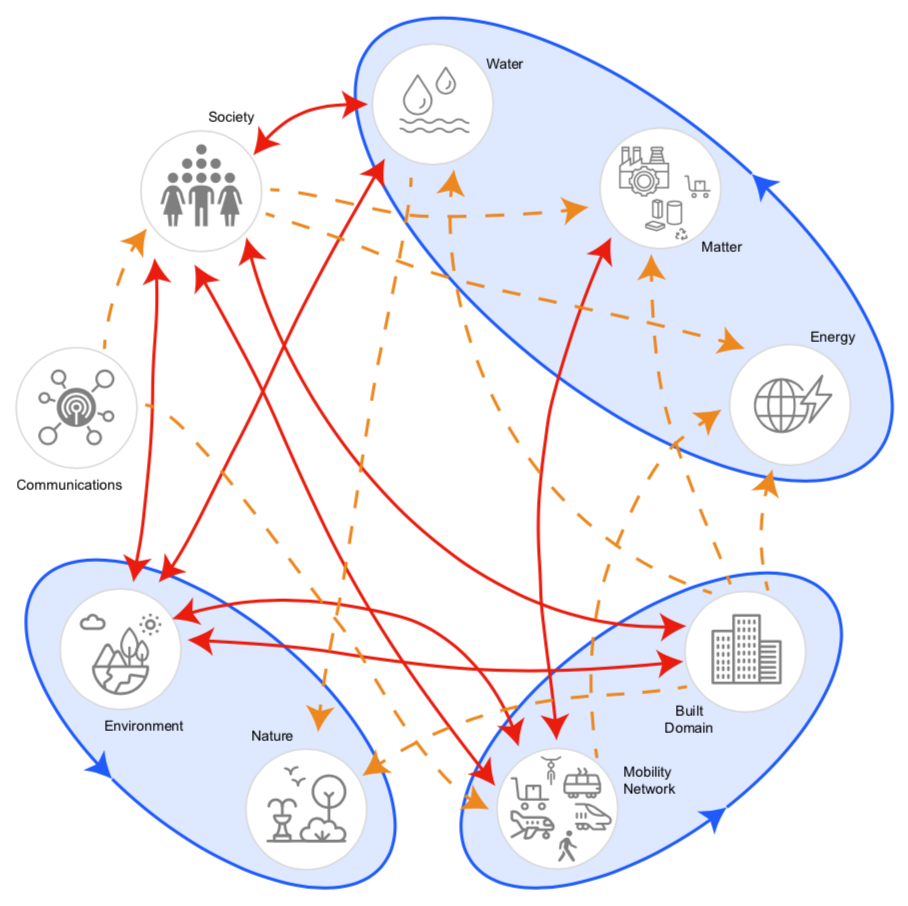}
\caption{{\bf City Anatomy -- Revisited. Towards City
Physiology.}
White circles: urban systems; blue ovals: strong cohesion; \colorA{red arrows:
double-way interactions; orange dashed arrows: one-way interactions. }
\colorA{Each orange arrow represents a cell of Fig. \ref{fig_connections_table}, where the direction points to the output system, from the input system. Each red arrows represents two cells, because the interaction is bidirectional.
Blue arrows don’t have a direction and are shaped as ovals, because the layers they enclose have a stronger interaction among them than with other systems. Created using icons from the \cite{icons_thenounproject} under a CC license.}}
\label{fig_anatomy_revisited}
\end{figure}

The table in Fig. \ref{fig_connections_table} helps in detecting strong interactions, especially when they are not a well-known feedback cycle, as in the case of Built Environment and Mobility Network. For instance (see orange cells of the table): to model the Mobility Network (= system of output), data from the Matter Cycle (= system of input) can be gathered about solid waste management policies and the routes followed by collectors. At the same time, to model the Matter Cycle (=system of output), routing data for
the transport and delivery of goods (Mobility network = system of input) may be useful. The direction of the loop depends on data availability and the purpose of the simulation.

An empty cell could show either a new research field (because the connection has never been explored before) or missing information (limited by the research criteria for references of `\nameref{city-sys-and-simu}' section). Both cases show possible next steps in developing the framework proposed in this paper. For instance, wireless communications (Communication Network) might be affected by building construction materials and shape (Built Environment), but the cell is empty because this specific research was not included in the review.

The original diagram of the City Anatomy organizes the systems according to a hierarchic order from macro to micro: `Structure', `Interactions' and `Society'. It was seen, however, that some systems have a strong cohesion between them 
\colorA{(e.g. WEF nexus)}, 
because they have a stronger interaction among them than with other systems. It was decided to revisit the diagram of the City Anatomy (Fig. \ref{fig_anatomy_revisited}) where not all the interactions are represented with the same intensity, but stronger interaction among system are evident.

\subsection{Interaction and data exchange
}\label{interaction-and-data-exchange}

Interaction patterns bring a technical problem to the development of the simulator; to run the simulation, the following steps need to be defined: \colorA{first, how to make each computational system progress each simulation-step (iterative or time, depending on the algorithm), then, which order the systems follow to iterate between each other.}

The interaction can be \colorA{either of convergence (the system reaches a point of equilibrium in a given time-step), or a temporal one (the interaction follows a sequence in time, with no feedback loops).}
Combinations or hierarchies of the two above mentioned options are also possible.

An interaction needs data exchange to occur in both in the \textit{integration of simulations} and \textit{the integration of software}. 
\colorA{
Simulation processes have different integration levels: the integration of simulation depends on whether every process of the simulator runs through all temporal series one after the other (sequential) or the simulator runs all processes in one temporal series before going to the next (parallel); the integration of software depends on whether all algorithms use the same coding language and are contents of the same software (supermodel) or they communicate within the same container (interface) or both (hybrid).
}
The same data might be required in different formats (e.g. land use is needed both for mobility and drainage simulations) and the output of every simulator needs to be feedable to other models as input (e.g. the result of a traffic simulation is the input for an emission simulator).

Data exchange can be structured with an ontology, where the relationships between content are defined and structured (e.g. \citet{DBpedia}). An example of ontologies related to urban models is the City Geography Markup Language (CityGML) \citep{CityGMLOGC}. See \nameref{SI_Appendix_ontologies} for a brief introduction to ontologies.

The City Anatomy Ontology of the CPS (\citeyear{cityprotocolsocietyCityAnatomyOntology2015}) is an attempt at a solution that formalises the relationships between its main elements and indicators, organising the terms and their relationship. However, the creation of an ontology is out of the scope of this paper, although interactions studied in this section could be a starting point for this purpose. The authors of this paper acknowledge the issue and leave it open for future work.

\subsection{Pipeline and practical example}
\label{pipeline}

\colorA{In this subsection, the framework envisioned in this paper is applied to develop integrated simulations, with the aim of solving a city what-if scenario. The steps to carry out the integration strategy are described, but technical implementation is left for future work.}
\colorA{The pipeline steps (see Fig. 3) 
are the following: problem definition, identification of layers involved and their interactions, identification of available simulators for each layer, and definition of coupling strategy.}

\begin{figure}[!ht]
\includegraphics[width=\textwidth]{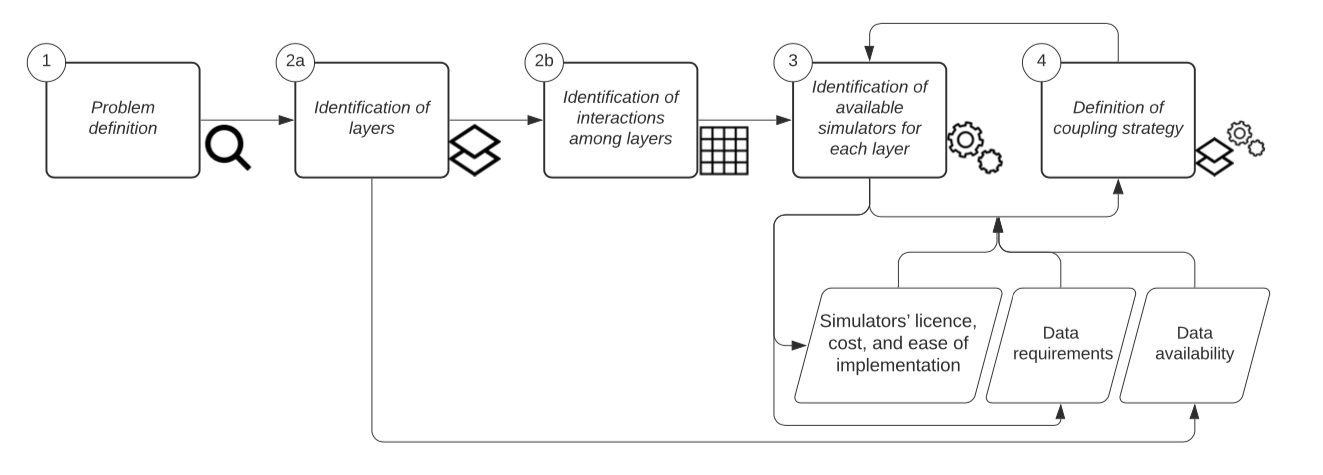}
\colorA{
\caption{{\bf Pipeline.}
Four pipeline steps:  problem definition, identification of layers involved and their interactions, identification of available simulators for each layer, and definition of coupling strategy. Steps 3-4 require iteration because they affect each other. }}
\label{fig_pipeline_steps}
\end{figure}

\colorA{As a practical example, we consider street pacification actions and their effects on people's health (problem definition). 
Street pacification has become a relevant action in cities as a non-pharmaceutical interventions in the context of contagious diseases. 
It involves changes in the distribution of traffic and on
the walkability of neighborhoods (Mobility Network), emission distribution (Environment - Air), and  effects  of  pollution  on  population  health  (Society).
We use Fig. \ref{fig_connections_table} and section \ref{system-crossing-interactions} (\nameref{system-crossing-interactions}) to define the interactions between these three layers. Starting from the column `Air' as system of output of Fig. \ref{fig_connections_table}, air quality simulations are affected not only by traffic emissions and environmental information such as climate and weather data, but also by the geometrical and thermal properties of the urban canopy (Built Domain) and the green canopy (Nature). The demographic data of the Society system is used to understand population mobility, both on a daily basis, transport use, and demand. Additionally, specific sectors of the population considered to be a sensitive population can be identified. If demographic data is not available, the Communication Network can provide location data and semantic information from digital traces.}

\colorA{We can use Section \ref{city-sys-and-simu} of this article, which contains a description of simulators organized by city layers, to identify which are the available simulators for each identified layer in the problem. However, further and more specific research should be done before moving to the next step of the pipeline, including other issues that influence the choice of simulator such as data availability and requirements, and the simulators’ licence, cost, and ease of implementation. It is important to underline that the last steps of the pipeline, where simulators are identified and the coupling process is defined, require iteration because they affect each other. 
Simulators with the same subject can be compared on the basis of characteristics such as spatio-temporal resolution, data requirements and modelling technique. While subject, data requirements, and outputs of the simulation highlight the potential data exchange between different city layers. A basic typology to compare models and simulators is suggested in \nameref{SI_Appendix_models}.
A possible development of the practical example consists of choosing an agent-based model to synthetically simulate the Society layer, and its activity schedule. The travel choice made by agents influences the travel paths to be followed, and their position in time and place is calculated. Pollutant emission is then calculated from vehicle position, and its propagation is computed with fluid dynamic simulations, with gradients refined at finer scale considering the influence of microclimate. The gradients are then compared with the position of agents. Depending on the choice of simulators and their programmatic interfaces, coupling between them can be achieved through standard software communication architectures (e.g. \citet{kuhnsimulatorcoupling}).} 

\colorA{A good example of air pollution and impact modelling is the second to last framework of \nameref{SI_Table_typology} (the same chosen in Table 3 of \nameref{SI_Appendix_models}), where submodels requirements and interactions are described. \nameref{SI_Table_typology} describe a case-study for every city layer, following the typology of \nameref{SI_Appendix_models}.}


\section{Roadmap and future work}
\label{roadmap-and-future-work}

Simulating the functioning of a city allows urban planners to evaluate the effects of multiple alternatives before implementing any relevant investment, as well as to detect potential urban disfunctions. 
The city is the core study of numerous varied disciplines which consider different specific aspects or activities. This article strives to collect together and organize these disciplines (see `\nameref{city-sys-and-simu}' section) under the structure of the City Anatomy. 
This being an overview of urban systems, the logic it was based upon allows other disciplines to be added, either new or existing (this could be the subject of further investigation and it is not dealt with in this paper).

\colorA{This overview allows the understanding of the data exchange between city models and the comprehension of the interactions between city systems (see `\nameref{developing-int-multisim}').
Following the organization of the `\nameref{city-sys-and-simu}' section, seven case studies were chosen and detailed in \nameref{SI_Table_typology}, that is published as supporting information  and in an open repository \citep{irem_im_2020_4014187}, so that new extended versions can be uploaded with further contribution from the research community.}

The outcome of this research is \colorA{to provide a framework} to define the elements and interactions of a holistic city simulator. Indeed, as the CPS (\citeyear{cityprotocolsocietyCityAnatomyFramework2015}) does not explain in detail how each system interacts with the others, the outcomes of this research were achieved with the help of \colorA{the results shown in Fig. \ref{fig_connections_table}.}
In the `\nameref{developing-int-multisim}' section the interactions between urban systems found in the existing literature have been listed.

The tool developed with this article is a starting point to a holistic approach to simulating urban systems --- more specifically, criteria to find what is already done, to understand how the systems work, and to understand the connections between systems are suggested and used. Ideally, other researchers will use this framework for simulating different aspects of a city in an integrated approach, and they will also contribute to it.

In the `\nameref{city-sys-and-simu}' section, it can be observed that urban metabolism models generally have a good holistic vision, especially within air and water models, which have reached an interesting level of integration and an in-depth detail of simulation. The scale of the model is constantly changing from field to field and it is evident that global --- or meso --- scales are needed to properly describe the environmental behaviours (e.g. air quality and water drainage) that affect cities.

This study provides a qualitative review of simulations from different research fields, that can be organized according to their contribution to urban systems, as well as a framework to study the interaction between urban systems and spot new research lines opened by the appearance of previously unseen connections.
\colorA{With a different approach, this framework shares the same purpose of urban digital twins, and hopefully it will help them in extending their domains to other city layers, as much as it will hopefully help urban practitioners in developing or engaging with simulators. In fact, interactions and coupling are not implemented, but the City Physiology defines a conceptual framework focused on the complexity that arises from the interaction among city layers.}
\colorB{Its methodology is independent of city and country by design. It can be applied to different countries because it is general enough not to be anchored to a specific context. Our modular approach ensures that each city or use case has the flexibility to configure a holistic simulator independently from others. This methodology is considered global because it considers every aspect of the city. Each city will have available data for some of the possible simulators, and can reuse simulators/data from other cities for the missing layers. Of course, urban settlements with different sizes behave differently, so the relationships between layers are different for a metropolis and a small village; still, the composition of layers remains the same, because the layers identified by the City Physiology describe the elements common to every urban settlement.}
However, several limitations of this study are acknowledged. 
First, the heterogeneity in key-concepts definition and keywords synonyms change from field to field.
This peculiarity of the area of study, that has been encountered since the definition of \textit{city}, could encourage the development of a unified terminology concerning the city.
Secondly, the heterogeneity in the research fields studied led us to consider reviews from only the last five years, when possible. So, this paper may not have individuated  relevant articles from previous years.
As the framework is being provided as an open tool, future work can be added, to increase its completeness.
Finally, the City Anatomy is not a peer reviewed publication, although it holds an ISO certification.

\begin{figure}[!ht]
\includegraphics[width=\textwidth]{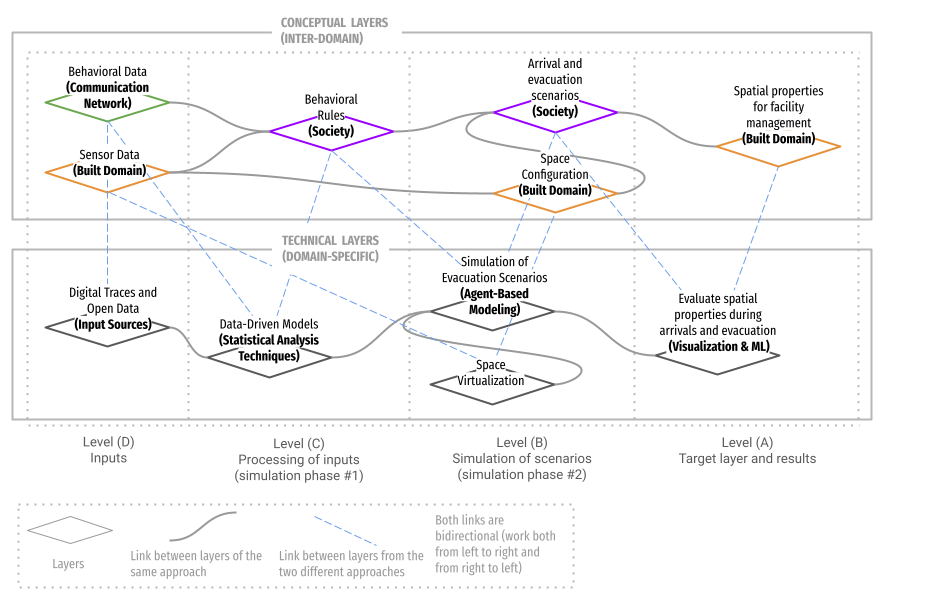}
\colorB{ 
\caption{{\bf Case study conceptual and technical layer.}
The figure, taken from \citet{CampNouTestBed}, shows how technical and conceptual layers assemble into the inter-domain and the domain-specific approaches. The dashed lines show the link between conceptual and technical layers: the conceptual layers are the output of the process defined in the technical layers.}}
\label{fig_IoTwins}
\end{figure}

Future key steps would be: first to extend the framework, adding case studies to exhaustively name all the elements of each system and how each system works and interacts with the others. Fig.\ref{fig_connections_table} is the starting point to identify new fields to progress in the line of research. After classification, the next step would be to develop simulations exploiting the interactions previously identified.
To do so, data conversion and understanding needs to be addressed.
\colorA{In the `\nameref{pipeline}' section the steps to apply the framework to an example are briefly introduced, but the implementation is out of the scope of this article and is left as future work. The next articles are planned to extend the pipeline steps into a complete methodology, and to apply it as a proof of concept.} 
\colorB{In the first follow-up article \citep{CampNouTestBed}, City Physiology is successfully tested. As depicted in Fig. \ref{fig_IoTwins}, that is taken from the published article, the framework is used to organise the simulation step from both a a conceptual and technical point of view. The conceptual layers are the output of the process defined in the technical layers; having the framework set up as so allowed the different experts involved in the simulation (i.e. technicians and managers) to use a common language that molded technical steps on policy-making goals.} \colorA{Technical problems are only hinted in the `\nameref{interaction-and-data-exchange}' subsection, but an \textit{interaction of software} through \textit{interface} (rather than a \textit{supermodel}) is a divide-and-conquer approach that allows modularity, extensibility, and flexibility. Being the simulations integrated through a container, it allows to focus on the simulation of one layer, which can than be connected to simulations of other layers.}

If these results are useful in identifying interactions within and between elements, other researchers, urban practitioners, and policy-makers will hopefully adopt, critique, and contribute to this approach. 
All tools in this research are offered as open repositories.
The authors of this paper believe that this open and public attitude should be adopted by the cities of the future in order to integrate simulators within their administrative structures as key tools for their management and administration. 
\colorA{The framework, in fact, provide a method and a vocabulary to develop or engage simulators.}
Indeed, holding data and information means having understanding and further control over the events that have generated them, and thus govern the city --- e.g. the real estate registry office is a control tool with which public administration exercises power. It is therefore proposed by the authors that the development of a powerful full city simulator should not be directed by private companies but by cities and public administrations instead.

According to the metaphor of the city as a body, the contribution of this paper could be framed as a \textit{City Physiology}. The CPS (\citeyear{cityprotocolsocietyCityAnatomyFramework2015}) provides what they call an `anatomy': the hierarchical relationship between systems. Instead, our contribution seeks a functional definition of the city systems: how interactions take place and how information moves from one layer to another, and, eventually, how the entire system works as a whole.

\section{Appendices and supporting information} 

\paragraph*{Table A}
\label{SI_Table_modeldefinition}
{\bf Modelling and models definition.} Additional table about urban-related definitions. 

\paragraph*{Table B}
\label{SI_Table_typology}
{\bf Typology table.} Seven case studies are classified following the typology \blue{developed in the \nameref{SI_Appendix_models}}. An editable version is available at \citep{irem_im_2020_4014187}.

\paragraph*{Appendix C}
\label{SI_Appendix_anatomy}
{\bf City Anatomy.} This appendix describes the systems considered by the City Protocol Society. 

\blue{
\paragraph*{Appendix D}
\label{SI_Appendix_models}
{\bf Classifying urban models.} This appendix provide a common glossary and suggest a typology to categorize simulation models. 
}

\paragraph*{Appendix E}
\label{SI_Appendix_ontologies}
{\bf Ontologies.} This appendix briefly introduces ontologies, in continuity with the `\nameref{interaction-and-data-exchange}' subsection. 

\bibliography{mybibfile}

\clearpage
\appendix

\section{Table: Modelling and models definition}

\begin{table}[ht]
\footnotesize
\centering
\caption{{\bf Definitions of urban concepts described in `Urban modelling' subsection.}}
\begin{tabular}{|p{3cm}|p{8.5cm}|p{2cm}|p{1cm}|}\hline
\bf Urban concept & \bf Definition & \bf Author & \bf Year
\\ \thickhline
Urban Planning Models 
& `...models which have been designed for, or
used in, urban physical and economic planning.' \cite{kilbridgeConceptualFrameworkUrban1969}
& Kilbridge et al. 
& 1969 
\\ \hline
LUTI - Land use Transport Interaction 
& `These early models were
equilibrium-seeking rather than dynamic, aggregate at the level of
populations involving spatial inter- actions, and built on conceptions
of the city articulated using ideas from urban economics and social
physics. They are usually now referred to as Land Use Transportation
Interaction (LUTI) models.' \cite{battyGenericFrameworkComputational2012}
& Batty 
& 2012 
\\ \hline
Large scale urban models (LSUMs) 
& `The term "large-scale model" is not
precise. "Large" is relative, changing with, among other things, the
power of computers. The urban models of interest here are those that
seek to describe, in a functional/structural form, an entire urban area,
portrayed in spatial, land-use, demographic and economic terms. Spatial
disaggregation yields zones that number at least in the hundreds, and
maybe the thousands.' \cite{leeRetrospectiveLargeScaleUrban1994}
& Lee 
& 1994
\\ \hline
Urban Models 
& `Representations of functions and processes which
generate urban spatial structure in terms of land use, population,
employment, and transportation, usually embodied in computer programs
that enable location theories to be tested against data and predictions
of future locational patterns to be generated.' \cite{battyUrbanModeling2009}
& Batty 
& 2009
\\ \hline
Integrated urban models (IUMs) (aka, integrated transport/land-use
models) 
& `Going back to the dawn of land-use modeling in the 1960s,
the initial motivation for developing such models was to generate the
population and employment inputs required by regional travel demand
models. Beginning in the 1970s the importance of understanding the
impact of major transportation investments (particularly transit) on
land values was increasingly recognized, as well as the importance of
land value impacts on benefit-cost justifications of major (and
expensive) transportation infrastructure (again, particularly
transit).' \cite{millerIntegratedUrbanModeling2018}
& Miller 
& 2018
\\ \hline
Urban Dynamics 
& `Representations of changes in urban spatial structure
through time which embody a myriad of processes at work in cities on
different, but often interlocking, time scales ranging from life cycle
effects in buildings and populations to movements over space and time as
reflected in spatial interactions.' \cite{battyUrbanModeling2009}
& Batty 
& 2009 
\\ \hline
Model
& `Our working definition of the term model is somewhat narrow:
the symbolic representation of urban relationships. Thus a model need
not derive directly from theory, but it must abstract urban phenomena to
symbolic form and relate these in a structural and mathematically
operational way.' \cite{kilbridgeConceptualFrameworkUrban1969}
& Kilbridge et al.
& 1969
\\ \hline
Urban Modelling 
& `The process of identifying appropriate theory,
translating this into a mathematical or formal model, developing
relevant computer programs, and then confronting the model with data so
that it might be calibrated, validated, and verified prior to its use in
prediction.' \cite{battyUrbanModeling2009}
& Batty 
&2009
\\ \hline

\end{tabular}
Definitions follow the same order as in the subsection.
\label{table_definitions}
\end{table}

\clearpage 

\section{Table: Typology table}

\includegraphics[width=\textwidth]{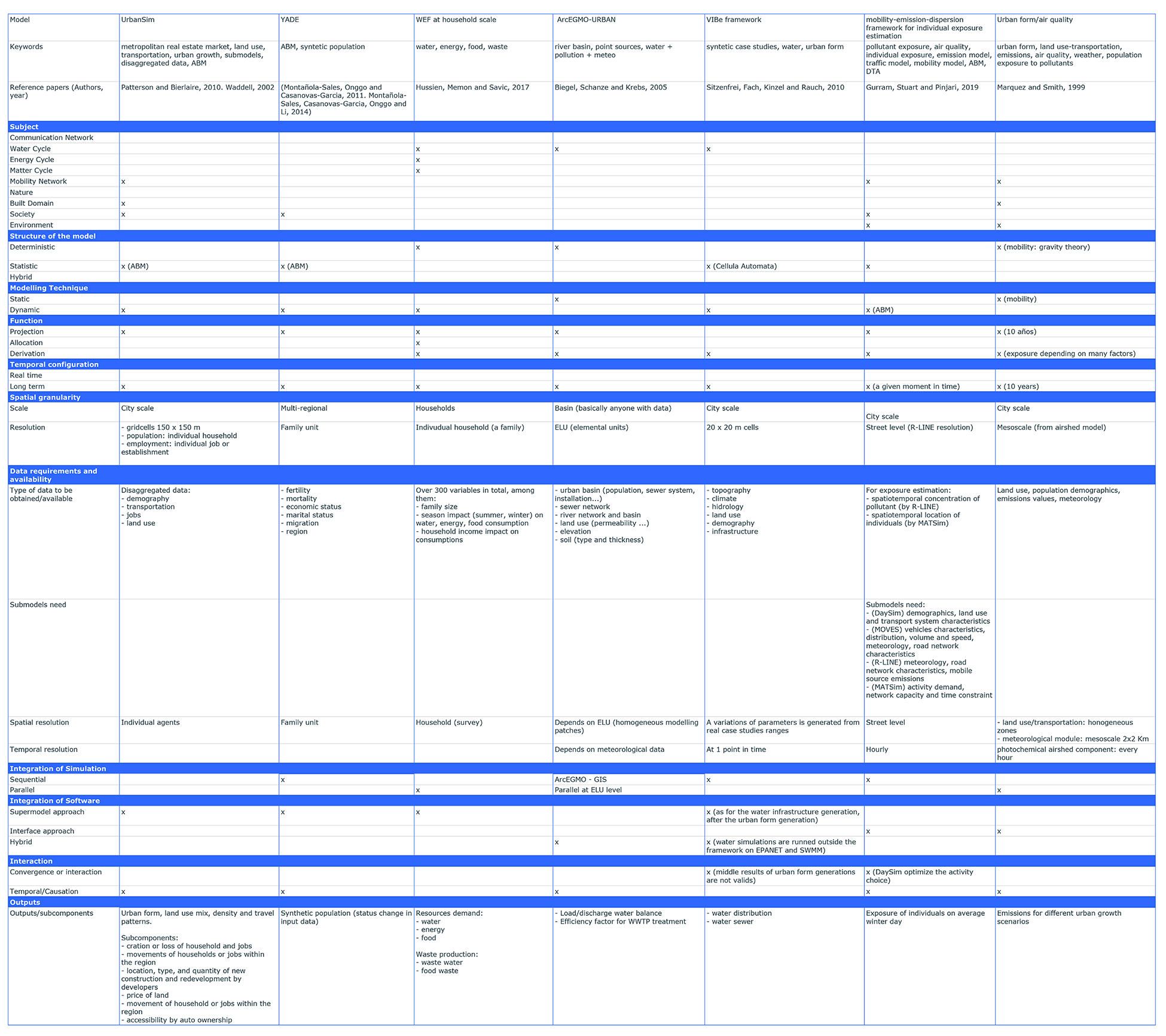}

\clearpage
\section{City Anatomy}

To start the research, the reference definition considered is the ‘anatomy’ given by  the \citet{cityprotocolsocietyCityAnatomyFramework2015}, that holds a ISO certification \citep{isoISO3710520192019}. The systems considered as basic and common to every city are:
\begin{enumerate}
    \item \textit{Communication Network}: it’s the information infrastructure and it’s made of both analogical elements and digital ones. Communication models can be centralized ones (like radio or television) with one emitter and many receivers, or distributed ones (from many to many).
    \item \textit{Water Cycle}: this infrastructure comprehends all the physical elements that work structurally to manage water (both clean and wastewater) in the city. It includes supply, depuration and sanitation, management and distribution of clean water, wastewater and surface water, and the draining and collection of the latter one.
    \item \textit{Energy Cycle}: this infrastructure manages the energy distribution towards and within the city, at all scales. It consists of functional nodes (nuclear plants, biomass, hydroelectric and solar plants, and wind farms), of a distribution network for energy (gas or electric one) and of a transport system for combustibles and chemical products (raw or refine ones). The scale of the infrastructure ranges from big nodes of production outside the city to small household consumption nodes. Different distributed models can be implemented within the city for the integrated energy production and management (photovoltaic panels or solar thermal collectors).
	\item \textit{Matter Cycle}: it’s the infrastructure involved in the extraction of all material resources necessary to the city (food included), in the matter manipulation (on both small and industrial scale) into consumable products, in the logistic and transport services till the consumers, in the consumption and management of waste (transport to dumps or recycling and energy or new materials production). There are two groups of materials: (a) all the matter used in consumption products and construction, and (b) the food, from animals or vegetables. Food is particularly important because, being almost a third of the annually world production lost in waste, it has a big impact onto the financial and environmental spheres, and onto the waste production.
	\item \textit{Mobility Network}: this infrastructure is related with transportation, mainly human but also of matter. It’s made of big-scale systems (airports, railroads and motorways) and small-scale ones (streets and squares, also theater of social interactions among dwellers).
	\item \textit{Nature}: this green infrastructure is made of all the natural elements structurally organized within the city. It comprehends all non-human alive entities at any scale (flora and fauna). It’s presence within the city affects on city life quality.
	\item \textit{Built Domain}: the ensemble of all the elements – private and public ones – surrounding the public space. Intrinsically multiscale – the object, the building, the city – it’s the material expression of the city culture – because it’s built by the man. Together with land use, the built domain establishes where the basic functions of human life take place.
    \item \textit{Society}: the system made of (i) citizen and (ii) government. The citizen are all the people within the city borders – even sporadically. Citizen are individuals, but they can gather together according to preferences. The government is a portion of the society elected to serve the community.
    \item \textit{Environment}: the physical setting in existence before cities were built. It is formed by nature (plants and animals outside the city borders) and air, soil and water, interacting dynamically in a seasonally dependable way.
\end{enumerate}

\clearpage
\section{Classifying urban models} 

In this appendix the literature from section \ref{city-sys-and-simu} `\nameref{city-sys-and-simu}' of the main article is organized and classified:
in the following section the discussion is limited to spot the components and limitations of each model (\textit{deterministic} vs. \textit{statistical}), while in the methodology section (`\nameref{proposed-typology}') the categories and the categorizing method are explained in detail.

\subsection{Defining modelling and models}\label{defining-modelling-and-models}

Words like \textit{models} and \textit{simulation} are often used with different meanings. Hence, it is necessary to provide a common glossary. 
Both \textit{deterministic} and \textit{statistical} models are considered, while virtual three-dimensional models that offer only a visualization of the spatial distribution are excluded. As defined by \citet{kilbridgeConceptualFrameworkUrban1969}, simulation models are techniques producing `conditional forecasts' : predictions that are the projection of a group of relevant variables, according to the change in others variables. To simulate something means to create a mathematical model to describe it.

According to \citet{kilbridgeConceptualFrameworkUrban1969}, the basic elements common to all models are:

\begin{itemize}
\item
  the subject: the theme or entity;
\item
  the function (what the model does): to project (to estimate the future
  state), to allocate (to study the distribution of the subject among
  subclasses of use or demand in a precise moment), or to derive new
  subjects (to transform or to derive another subject from the original,
  like traffic from land use);
\item
  the theory: the set of rules and relationships (implied or explicit)
  that prevail between subject and the environment, on which the model
  works.
\item
  the method: the way the theory is used within the model and the
  techniques used to process the information (see \textit{analytic} and
  \textit{simulation models} in the paper).
\end{itemize}

Deterministic models are `theory-based' \citep{kilbridgeConceptualFrameworkUrban1969} models, defined by a series of equations that exist, are known, and can be solved or approximated through mathematical operations. These models rely on generally-agreed theories (see `secure' models by \citet{bullockLevinsLureArtificial2014}) or
at least a field of equations defined by experts explaining how the studied phenomenon works. Deterministic models can explain the reason behind an event. Furthermore, the relationship between theory and model is recursive: models can test the theories they are based on, and can help in defining a better one \cite{kilbridgeConceptualFrameworkUrban1969}.

Statistical models are \textit{black-box} or `theory-laden' \citep{kilbridgeConceptualFrameworkUrban1969} models, defined by a series of statistical equations (the model includes probabilistic relationships between elements). The output of the model can also be a probability, or at least it is a number extracted from a probabilistic distribution. Statistical models find patterns in data, much harder to parametrize by a human being, who does not have the ability to express such complicated rules (see the `exploratory attempts' of `insecure' models in \cite{bullockLevinsLureArtificial2014}). In these types of model, their construction and their results are more important than the behaviour of the simulation algorithms, that may be not fully understood \citep{bullockLevinsLureArtificial2014}. Understanding the meaning of the solution and how it will respond to changes in the input variables is a difficult process; even though the results are still concrete and falsifiable \citep{bullockLevinsLureArtificial2014}.

The law of `downhill design and uphill analysis' of Braitenberg \citep{bullockLevinsLureArtificial2014} subdivides the concept of `tractability' (that is, the ability to use the model productively) into tractability of component and tractability of comprehension.
The two are interdependent in deterministic models. In contrast, in statistical models, the effort of tractability is focused on the model design phase: the importance of the construction of components outweighs the understanding of the model behaviour.

According to \citet{bullockLevinsLureArtificial2014} `useful models' are a combination of `tractability' (compulsory) with `realism', `precision' and `generality'. It is not possible to maximize them all at the same time: a model can fulfil only two at the expense of the third one. Indeed, a model that combines realism, precision, and generality cannot be solved in practice: it cannot be compared with reality because it gives no results (see `functional model' by Barandian and Moreno in \citep{bullockLevinsLureArtificial2014}).
As \citet{bullockLevinsLureArtificial2014} explains in the title itself (`...the lure of artificial worlds'), the author highlights the danger of betting on maximising generality, precision, and realism at the same time. See Table \ref{table_det_VS_stat} for a synthetic comparison between deterministic and statistical models, and Table \ref{table_model_categories} for models' categories.

\subsection{Proposed typology }\label{proposed-typology}

A method to classify models concerning the urban context is proposed. The definition of categories is based on \citep{kilbridgeConceptualFrameworkUrban1969, berling-wolffModelingUrbanLandscape2004}, and on model features in \cite[Table 3, page 97]{bachCriticalReviewIntegrated2014}.
The categories are:

\begin{itemize}
\item
  \textit{Subject}: one or more layers of the `City Anatomy' \citep{cityprotocolsocietyCityAnatomyFramework2015};
\item
  \textit{Structure of the model}: deterministic (the theory behind the model)
  vs. statistic (black box/without theory), hybrid models are a
  possibility;
\item
  \textit{Modelling technique}: static vs. dynamic (ABM, CA, etc. are dynamic
  techniques);
\item
  \textit{Function}: the purpose of the model (to project, allocate or derive);
\item
  \textit{Temporal configuration}: real time vs. long term (results are used for
  immediate monitoring and control vs. results which predict possible futures
  or alternative pasts);
\item
  \textit{Spatial granularity}: scale and resolution of the model;
\item
  \textit{Data requirements and availability}: type of data to be
  obtained/available, their resolution and their reliability;
\item
  \textit{Integration}: if it's the case, integration level of processes;
  
  \begin{itemize}
  \item
    \textit{of simulation}: every process of the model runs through all temporal
    series one after the other (sequential) vs. the model runs all
    processes in one temporal series before going to the next
    (parallel);
  \item
    \textit{of software}: all algorithms use the same coding language and are
    contents of the same software (supermodel) vs. they communicate
    within the same container (interface) vs. both (hybrid);
  \end{itemize}
  
\item
  \textit{Interaction}: convergence vs. temporal sequence
  (causation);
\end{itemize}

\begin{itemize}
\item
  \textit{Outputs}: the outputs of the model or of its subcomponents.
\end{itemize}

The design of a model faces obstacles and barriers depending on model complexity, ease of use, administrative fragmentation and communication.
Its construction follows the basic steps of calibration, verification, and validation, that are based on the scientific method to develop a theory. From the first stages, parameters have to be calibrated (adapt real situation data to the particular situation), the internal consistency of the model has to be verified and the model has to be validated with data (to be sure it reproduces the phenomena of interest acceptably) \citep{bachCriticalReviewIntegrated2014, battyUrbanModeling2009, battySciencePlanningTheory2017}.

The basic steps of modelling are \citep{bachCriticalReviewIntegrated2014}:

\begin{itemize}
\item
  collecting data;
\item
  identifying the purpose of the simulation;
\item
  defining data requirements (data availability and typology affect the
  model's characteristics; besides, they influence calibration,
  verification and validation process);
\item
  data cleaning and selection of relevant variables;
\item
  (if integrated models) identifying of relevant interaction among
  sus-systems;
\item
  (if integrated models) selection of a integration method: by code or
  by interface;
\item
  defining the model(s);
\item
  calibration and validation;
\item
  decreasing, if possible, uncertainty (along the entire process);
\item
  outputs analysis;
\item
  remodelling, if necessary.
\end{itemize}

The typology explained above enables us to identify open problems in
the city modelling field;
in fact, the categories facilitate the comparison between models and the understanding of their potential coupling. Models with the same subject can be compared on the basis of characteristics such as spatio-temporal resolution, data requirements and modelling technique. At the same time, models that apply the same modelling technique to different subjects can be spotted. The categories of subject, data requirements and availability, and outputs highlight data exchange between different city layers (or submodels, if it is the case).

\subsection{Exemplification of the typology
}\label{exemplification-of-the-typology}

Following the reasoning in the previous subsection (`\nameref{proposed-typology}'), the cases studied in this research are classified in  \nameref{SI_Table_typology} and available at \citep{irem_im_2020_4014187}.
A meaningful example for the Environment layer is given in Table \ref{table_air_example}. 

Briefly, the chosen case study is an agent-based model developed by \cite{gurramAgentbasedModelingEstimate2019}; it was designed to estimate exposure to the air emissions of transportation and is a good example of integration and complexity. An agent-based technique is used to evaluate exposure to NOx from transportation for different transportation user groups in the Tampa area, Florida, USA. 
\begin{table}[!ht] 
\centering
\footnotesize

\begin{tabular}{|p{8cm}|p{7.5cm}|}
\hline
\bf Deterministic models & \bf Statistical models
\\ \thickhline
Field of known equations solved through mathematical operations 
& Probabilistic distribution between elements
\\ \hline
Theory-based 
& `Black-box': unknown internal relationships
\\ \hline
Prediction + Meaning of the solution 
& Prediction
\\ \hline
Tractability of components (components' design) 
& Tractability of components
\\ \hline
Tractability of comprehension (model behaviour) 
& Low tractability of
comprehension
\\ \hline
``Reverse'' process of going from the results of the simulation to an
informative solution (why that solution) 
& The understanding of the
algoritms' behaviour is not as relevant as the result
obtained
\\ \hline

\end{tabular}
\caption{
{\bf Deterministic and Statistical models in a nutshell.}}
\label{table_det_VS_stat}
\end{table}

\begin{table}[!ht]
\label{table_model_categories}
\footnotesize
\centering

\begin{tabular}{|p{8cm}|p{4cm}|p{2cm}|}
\hline
\bf Category & \bf Author & \bf Citation
\\ \thickhline
theory-based & Kilbridge & \cite{kilbridgeConceptualFrameworkUrban1969} \\ \hline
theory-laden & Kilbridge &  \cite{kilbridgeConceptualFrameworkUrban1969} \\ \hline
Simulation models & Kilbridge &  \cite{kilbridgeConceptualFrameworkUrban1969} \\ \hline
Analytic models & Kilbridge &  \cite{kilbridgeConceptualFrameworkUrban1969} \\ \hline
Secure models & Bullock &  \cite{bullockLevinsLureArtificial2014} \\ \hline
Insecure models & Bullock &  \cite{bullockLevinsLureArtificial2014} \\ \hline
Realism & Levins &  \cite{bullockLevinsLureArtificial2014} \\ \hline
Precision & Levins &  \cite{bullockLevinsLureArtificial2014} \\ \hline
Generality & Levins &  \cite{bullockLevinsLureArtificial2014} \\ \hline
Tractability & Levins &  \cite{bullockLevinsLureArtificial2014} \\ \hline
Mechanistic models (realism, precision, tractability) & Barandiaran y Moreno &  \cite{bullockLevinsLureArtificial2014} \\ \hline
Generic models (generality, precision, tractability)
& Barandiaran y Moreno
&  \cite{bullockLevinsLureArtificial2014} \\ \hline
Conceptual models (realism, generality, tractability)
& Barandiaran y Moreno 
&  \cite{bullockLevinsLureArtificial2014} \\ \hline
Functional models (realism, precision, generality) 
& Barandiaran y Moreno
&  \cite{bullockLevinsLureArtificial2014} \\ \hline
Mathematical (Equational) models & Peck &  \cite{bullockLevinsLureArtificial2014} \\ \hline
Simulation models & Peck &  \cite{bullockLevinsLureArtificial2014} \\ \hline

\end{tabular}
\caption{
{\bf Categories of models, as found in the reviewed literature.}}
\end{table}

\begin{table}
\footnotesize
\centering
\begin{tabular}{|p{5cm} + p{11cm}|}
\hline
\textbf{Subject} & Environment (air), Mobility Network \\ \hline
\textbf{Structure of the model} & statistic \\ \hline
\textbf{Modelling technique} & dynamic (ABM) \\ \hline
\textbf{Function} & projection, derivation \\ \hline
\textbf{Temporal configuration} & long term \\ \hline
\multicolumn{2}{|l|}{\textbf{Spatial granularity}} \\ \hline
Scale & city scale \\ \hline
Resolution & street level (R-LINE resolution) \\ \hline
\multicolumn{2}{|l|}{\textbf{Data requirements and availability}} \\ \hline

Type of data to be obtained/available
& \begin{minipage}[t]{0.9\columnwidth}\raggedright
for exposure estimation:
\begin{itemize}
\item
  spatiotemporal concentration of pollutant (by R-LINE)
\item
  spatiotemporal location of individuals (by MATSim)
\end{itemize}
\end{minipage}\\ \hline

Submodels need 
& \begin{minipage}[t]{0.8\columnwidth}\raggedright
\begin{itemize}
\item
  (DaySim) demographics, land use and transport system characteristics
\item
  (MOVES) vehicles characteristics, distribution, volume and speed, \\
  meteorology, road network characteristics 
\item
  (R-LINE) meteorology, road network characteristics, mobile source
  emissions
\item
  (MATSim) activity demand, network capacity and time constraint
\end{itemize}
\end{minipage}\\ \hline
Spatial resolution & street level \\ \hline
Temporal resolution & hourly \\ \hline
\multicolumn{2}{|l|}{\textbf{Integration}} \\ \hline
of simulation & sequential \\ \hline
of software & interface approach \\ \hline
\textbf{Interaction} & temporal/causation (supermodel) and convergence (DaySim
optimize the activity choice) \\ \hline
\textbf{Outputs} & exposure of individuals on average winter day \\ \hline

\end{tabular}
\caption{
{\bf Typology table applied to the environment case study.}}
\label{table_air_example}
\end{table}

\clearpage 

\section*{Appendix E: Ontologies}

This appendix briefly introduces ontologies, in continuity with the `Interaction and data exchange' subsection of the main article.

\textit{Ontology} comes from philosophy and it consists of studying the nature and structure of reality. In computer science, ontologies are `computational artifacts' \citep{yangOntologybasedSystemsEngineering2019}, that are `sharable and reusable knowledge repositories'\citep{benslimaneDefinitionGenericMultilayered2000}. That is: the `nature and structure' of the portion of reality taken into consideration (the `domain') is defined, based on communication, agreement, sharing and interoperability.

Ontologies are used in web technologies, database integration, multi-agents systems, and natural language processing \citep{rousseyIntroductionOntologiesOntology2011}.
They help to overcome obstacles to interoperability like \citep{benslimaneDefinitionGenericMultilayered2000}:
\begin{itemize}
    \item `conflicts from data models and types',
    \item `semantic discrepancies between different components' that are designed and used independently,
    \item `lack of tools to access shared information'.
\end{itemize}
Ontologies offer a solution to these obstacles because they \citep{benslimaneDefinitionGenericMultilayered2000}: 
\begin{itemize}
    \item `translate one information system to another system', 
    \item provide a common data format,
    \item 'use integration techniques to merge collections of information sources into federated databases'.
\end{itemize}
The modelling process consists of ad-hoc mapping of the real world into programming instances: this abstraction process can lead to inconsistencies between different users' points of view \citep{metralOntologiesInterconnectingUrban2011}. The consequent conflict can be avoided if the user's point of view is translated into an ontology, and then it's used for the development of the model \citep{metralOntologiesInterconnectingUrban2011}.

The main components of an ontology are: 
\begin{itemize}
    \item concepts --- or classes,
    \item instances --- or objects,
    \item properties --- or values. 
\end{itemize}
\citet{rousseyIntroductionOntologiesOntology2011} classify ontologies following three criteria: 
\begin{itemize}
    \item `based on language expressivity and formality',
    \item `based on the scope of the ontology, or on the domain granularity',
    \item `based on design approach'.
\end{itemize}

Following the first criterion, among `Linguistic terminological ontologies' fall Thesauri such as \citet{URBADOC} or the \citet{GEMETGeneralMultilingual}.
They focus on the listing of the vocabulary in use and its normalization (agreement on the meaning) \citep{rousseyIntroductionOntologiesOntology2011}.

Models can be interconnected through ontologies which allow data conversion and understanding \citep{benslimaneDefinitionGenericMultilayered2000}. First, the models need to be
represented through an ontology, then the different ontologies are interconnected \citep{metralOntologiesInterconnectingUrban2011}. An example of ontologies related to urban models is the City Geography Markup Language (CityGML) \citep{CityGMLOGC}.

The problems arising from interconnecting ontologies are quite similar to those from connecting models: 
\begin{itemize}
    \item different languages or tools used,
    \item setting up the detailed interconnection has to overcome the lack of biunivocal correspondence between concepts and the ambiguity in interpreting the terms,
    \item different scale or representation of the same object, that are not explicit.
\end{itemize}
Furthermore, the trade-off between constructing a large and extensive ontology inclusive of all concepts and relationships of a specific domain, and constructing a generic ontology with less concepts, but which a bigger community can agree on, must be considered \citep{benslimaneDefinitionGenericMultilayered2000}.

\end{document}